\documentclass{desyproc}

\begin{document}
\title{Axion-like particles and\\circular polarisation of active galactic nuclei}

\author{{\slshape A.~Payez, J.R.~Cudell, D.~Hutsem\'{e}kers}\\[1ex]
University of Li\`{e}ge, All\'{e}e du 6 Ao\^{u}t 17, 4000 Li\`{e}ge, Belgium\\}

\contribID{payez\_alexandre}

\desyproc{DESY-PROC-2009-05}
\acronym{Patras 2009} 

\maketitle

\begin{abstract}
	The measurements of the linear polarisation of visible light from quasars give strong evidence for large-scale coherent orientations of their polarisation vectors in some regions of the sky. We show that these observations can be explained by the mixing of the photons with very light pseudoscalar (axion-like) particles in extragalactic magnetic fields during their propagation. We present a new treatment in terms of wave packets and discuss the circular polarisation.
\end{abstract}

\section{Introduction}

	In this work~\cite{paper}, we are interested in the effect that axion-photon mixing can have on the polarisation of light coming from distant astronomical sources. In particular, the observations of redshift-dependent large-scale coherent orientations of AGN polarisation vectors can, at least qualitatively, even in very simple models, be reproduced as a result of such a mixing of incoming photons with extremely light axion-like particles in external magnetic fields. These observations, presented in the second edition of this conference, were based on good quality measurements of the linear polarisation for a sample of 355 measured quasars in visible light~\cite{hutsemekers}.
	
	This has been discussed in terms of axion-photon mixing by several authors, in the case of plane waves~\cite{planewaves} and a prediction from this mixing is an observable circular polarisation comparable to the linear one.
	Here, we present the case in which light is described by wave packets and show that the circular polarisation can be suppressed with respect to the plane wave case.

\section{Axion-photon mixing using Gaussian wave packets}

	\subsection{The idea behind this}

	The mixing of axion-like particles with photons is usually discussed mathematically in terms of infinite plane waves. Using that description, the Stokes parameters can be computed and predictions of the polarisation of light from the interaction can be given; the main properties of such a mixing being \emph{dichroism} and \emph{birefringence} (see~\cite{reviewHLPW} for a review of this case).
	While dichroism would be an interesting way to produce linear polarisation and, in particular, to explain the observations concerning quasars, birefringence ---which is linked to the creation of circular polarisation--- would give a very clear signature of the mixing. Indeed, in this formalism of plane waves, except in extremely specific cases, the \emph{circular polarisation} predicted \emph{can be as large as the linear polarisation}\footnote{This is what one obtains if one does not assume very specific distributions of magnetic field orientations along the line of sight.}.

	The idea discussed here is to send wave packets into a region of uniform magnetic field and to compute the Stokes parameters. Before the magnetic field, the wave packets have the form:
	\begin{equation}
		E (x,t) = \int_{\omega_p}^{\infty} \frac{d\omega}{N}  e^{-\frac{a^2}{4}(\omega-\omega0)^2} e^{i \sqrt{\omega^2 - \omega_p^2} (x-x_0)} e^{- i \omega (t-t_0)},\label{eq:packet}
	\end{equation}
	where $\omega_p$ is the plasma frequency of the medium and $a$ controls the initial width of the packet (in the limit $a\rightarrow +\infty$, this reduces to the plane wave case).

	The main motivation for considering this formalism comes from the measurements of circular polarisation of some of the quasars considered in~\cite{hutsemekers}. While axion-photon mixing would be an attractive explanation of the observations for linear polarisation, preliminary results show that circular polarisation of light from these AGN seems to be, in general, \emph{much smaller} than the linear polarisation~\cite{polcirc}. This means that if the creation of circular polarisation was really a smoking gun of ALP-photon mixing, no matter how refined the description, these observations would rule out the mixing mechanism and could only be used to constrain the existence of axion-like particles.

	For these reasons, it can be interesting to work with wave packets, as new effects will be taken into account, including dispersion, separation of packets and coherence; effects that might be of importance for the Stokes parameters.
	Note that calculating the propagation of packets of the form \eqref{eq:packet} is numerically\footnote{We use Multiple-Precision Floating-point library with correct Rounding: \url{www.mpfr.org}.} tricky, as the computation of the Stokes parameters requires a spatial resolution of the order of the width of the wave packets after a propagation over huge distances in the magnetic field (we will usually consider one magnetic field zone of 10~Mpc~\cite{vallee} and initial wave packets of width $\lesssim1\mu$m).

	\subsection{Results with wave packets}

		In the plane transverse to the direction of propagation, we choose a basis of two orthogonal linear polarisations, the same as the one used in the plane wave case, so that we will talk about polarisation parallel or perpendicular to the transverse external magnetic field $\vec{\mathcal{B}}$.
		This being done, we next choose the electric fields $E_{\parallel}(x,t)$ and $E_{\perp}(x,t)$ both initially described by a function of the form \eqref{eq:packet}. Then, we propagate these using the equations of motion for the electromagnetic field which take into account the interaction with pseudoscalar particles and find the expressions of the electric fields after a propagation, when axion-photon mixing is at work, inside a step-like magnetic field region.

		\begin{figure}
			\begin{center}
				\includegraphics[width=0.69\textwidth]{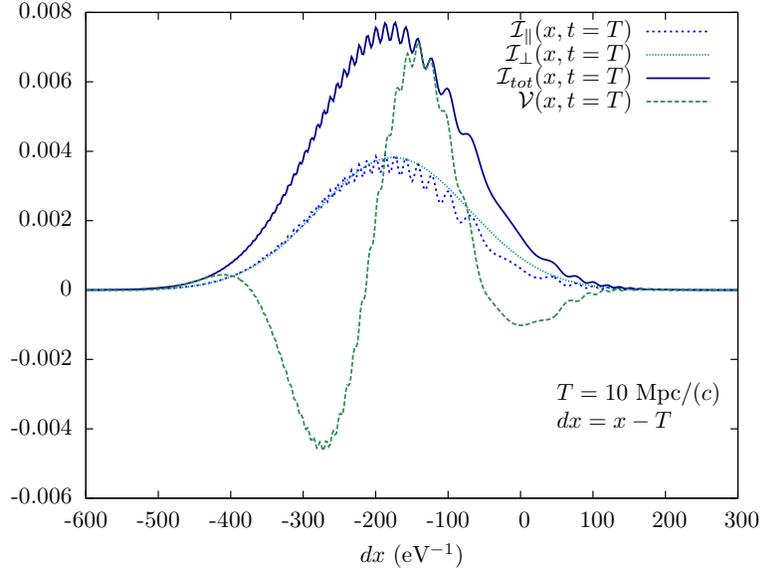}
				\caption{Wave packets: illustration after a propagation time $T$ in an external magnetic field, in a strong mixing case ---here, the axion mass is $m=4.7~10^{-14}$~eV, $\omega_p = 3.7~10^{-14}$~eV and $g\mathcal{B}=5.5~10^{-29}$~eV. The initial width of the wave packet has been chosen $\simeq\lambda_0$.}
				\label{fig:packets}
			\end{center}
		\end{figure}

		We can then use the expressions of the Stokes parameters ---which are observables built on intensities--- that can, for example, be plotted as functions of $x$, the distance travelled inside the magnetic field, for a given propagation time, $T$. This is what is represented in Figure~\ref{fig:packets} which shows what the two packets look like (respectively $\mathcal{I}_{\parallel}(x,t=T)$ and $\mathcal{I}_{\perp}(x,t=T)$) but also the total intensity (which is just the sum of the two) and the unnormalised circular polarisation, $\mathcal{V}(x,t=T)$.
		This is for a beam with a central wavelength $\lambda_0 = 500$~nm, initially 100\% linearly polarised, with its polarisation plane making a 45$^\circ$ angle initially with the magnetic field direction (i.e. $u(0) = \frac{U(0)}{I(0)}= 1$; $q(0) = v(0) = 0$)\footnote{$u$ and $q$ are the two Stokes parameters required to describe fully the linear polarisation of a light beam, while $v$ accounts for the circular polarisation.}; this angle being, in fact, the most favourable one for the creation of circular polarisation, due to birefringence.
	Note also that the abscissa is $dx$, the shift in position with respect to a frame moving a the speed of light $c$ (namely, a maximum at $dx = 0$ corresponds to $|\vec{v}|=c$).

		From the observational point of view, there is a macroscopic exposure time over which one should integrate these functions to obtain, finally, the value of the observable Stokes parameters, e.g.:

		\begin{equation}
			I(x) = \int_\mathrm{exposure~time} dt~\mathcal{I}(x,t)\nonumber.
		\end{equation}

\begin{figure}
			\begin{center}
				\includegraphics[width=0.69\textwidth]{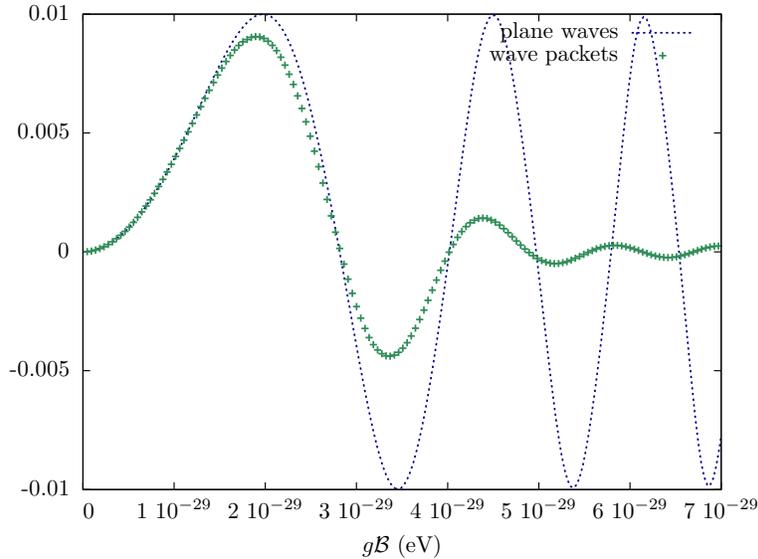}
				\caption{Circular polarisation for different values of the coupling in the case of initially partially linearly polarised light with $u(0)=0.01$. $\lambda_0$, $\omega_p$ ---and $a$, for the wave packet--- are the same as in Figure~\ref{fig:packets}, other parameters are: $T=10$~Mpc, $m=4.3~10^{-14}$~eV.
				}
				\label{fig:circpol}
			\end{center}
		\end{figure}

		From these integrals, we obtain that the wave packet formalism leads to a circular polarisation, $v=\frac{V}{I}$, \emph{lowered} with respect to plane wave case. Figure~\ref{fig:circpol} illustrates the plane wave ($a\rightarrow\infty$) result: it shows the amount of circular polarisation gained due to axion-photon mixing with different values of the coupling $g\mathcal{B}$ ($g$ being the axion-photon coupling constant). In that simpler case, it is known that $v=\frac{V}{I}$ oscillates between $-|u(0)|$ and $|u(0)|$, whereas in the wave packet case it is shown that there is a damping of these oscillations. It follows from this observation that \emph{$v$ is no longer expected to be as large as the linear polarisation in general}.

\section{Conclusion}

	We have briefly presented axion-photon mixing with the use of wave packets. The main consequence of this treatment is the net decrease of circular polarisation with respect to what is predicted using plane waves. Hence, the lack of circular polarisation in the light from AGN does not rule out the ALP-photon mixing.

\section*{Acknowledgements}

	A.~P. would like to thank the IISN for funding and to acknowledge constructive discussions on physical and numerical matters with Fredrik Sandin and Davide Mancusi.



	\begin{footnotesize}
		
	\end{footnotesize}


\end{document}